\documentclass[slac_one]{revtex4}
\usepackage{graphicx}
\DeclareGraphicsExtensions{.pdf,.eps}
\usepackage{fancyhdr}
\begin{document}

\title{Commissioning of the ATLAS reconstruction software with first data} 

%

\author{M. Moreno Llacer (for the ATLAS collaboration)}
\affiliation{IFIC (CSIC-UV), Valencia, Spain}

\begin{abstract}
Looking towards first LHC collisions, the ATLAS detector is being commissioned using the physics data available: cosmic rays and data
taken during the LHC single beam operations at 450 GeV. During the installation of the ATLAS detector in the cavern, cosmic rays were collected with the different parts of the detector that were available. Combined cosmic runs taken with the full installed detector with and without magnetic field as well as a few single beam events recently recorded are being used to commission the full
system prior to the first proton collisions.
\end{abstract}

\maketitle

\thispagestyle{fancy}


\section{Introduction}
\begin{figure}[h!]
\begin{center}
\includegraphics[width=0.6\textwidth]{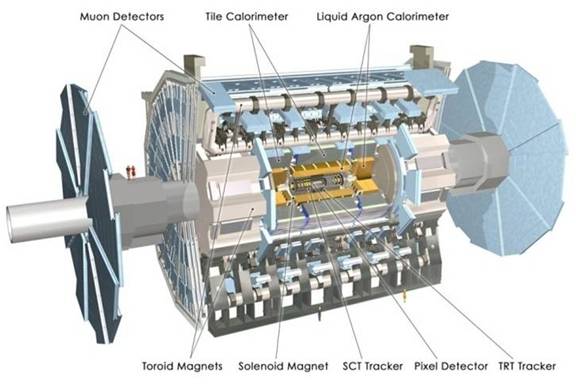}
\caption{\sl Schematic view of the ATLAS detector. The dimensions of the detector are 25 m in height and 44 m in length. The overall weight of the detector is approximately 7000 tons.}
\label{ATLAS}
\end{center}
\end{figure}

ATLAS (A Toroidal Lhc ApparatuS)~\cite{ATLAS} is one of the four major experiments at the forthcoming LHC (Large Hadron Collider), in which protons will collide at a center of mass energy of 14 TeV. It consists of three main sub-systems: the Inner Detector (ID), the Calorimetry system (electromagnetic and hadronic calorimeters) and the Muon Spectrometer (MS). It uses a superconducting magnet system with a central solenoid around the inner detector and large air-core toroid magnets for the muon spectrometer. Fig.~\ref{ATLAS} shows the overall detector layout.\\
The commissioning of the ATLAS detector with physics data already started while the detector was being mechanically and electrically completed by
collecting cosmic rays with the parts of the detectors that were becoming available. Global cosmic rays runs with the complete detector with
different magnetic field configurations are now being recorded. The full system was ready to collect data during the three days in which a single
beam of the LHC at the injection energy of 450 GeV circulated through ATLAS.\\
In addition to put in place the trigger and data acquisition chains, commissioning of the full software chain is a main goal. This is interesting not only to ensure that the reconstruction, monitoring and simulation chains are ready to deal with LHC collisions data, but also to understand the detector performance in view of achieving the physics requirements. Furthermore, they have been used to validate and improve the ATLAS simulation comparing the results obtained from real data and those from the specific simulations.\\

\section{Reconstruction chain}
\begin{figure}[h]
\begin{center}
\includegraphics[width=0.35\textwidth]{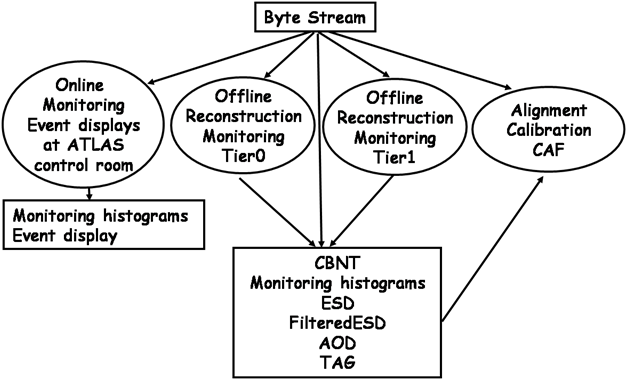}
\caption{\sl Schematic of the software chain.}
\label{SoftwareChain}
\end{center}
\end{figure}
\begin{figure}[!h]
\begin{center}
\includegraphics[width=0.4\textwidth]{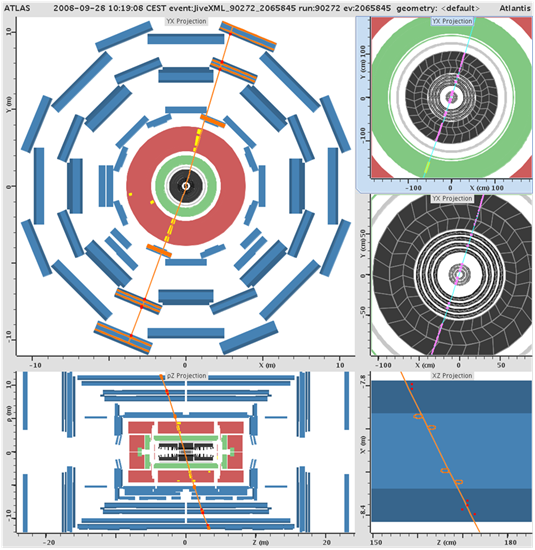}
\caption{\sl Event display showing a cosmic ray crossing the ATLAS barrel, recorded during ATLAS combined cosmic run 90272 which the full magnetic field. A combined (inner detector and muon spectrometer) track is reconstructed in this event.}
\label{ED2}
\end{center}
\end{figure}
\begin{figure}[h!]
\begin{center}
\resizebox{0.57\textwidth}{!}{
\includegraphics[width=0.27\textwidth]{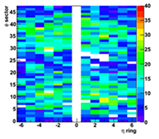}
\includegraphics[width=0.30\textwidth]{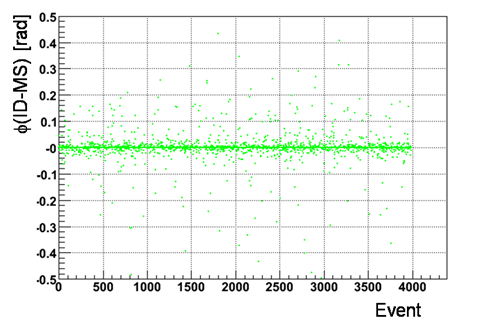}
}\\
\caption{\sl SCT modules occupancy (left). Difference of the $\theta$ track parameter measured by the inner detector and muon spectrometer as a function of the event number (right).}
\label{Monitoring}
\end{center}
\end{figure}

The commissioning period is being used to put in place the full detector operation chain, i.e. from the LVL1 trigger and data acquisition to the analysis in the Grid Tier-2 computing centers. The software plays an important role in this chain.\\
The software chain is shown in Fig.~\ref{SoftwareChain}. Cosmic rays and events recorded during LHC single beam operations are reconstructed using the full ATLAS software chain, with specific modifications to account for the lack of synchronization of these kind of events with the readout clock (even during single beam operations due to the not yet operational RF capture) and the fact that particles do not come from the center of the detector. The reconstruction and monitoring algorithms have been continuously running online in the ATLAS control room to provide online event displays and histograms monitoring the data quality during detector operations. Event displays in which a cosmic ray track is reconstructed is shown in Fig.~\ref{ED2}. One can see hits in the trigger and precision muon spectrometer chambers, energy deposited in the calorimeters and hits in the inner detector. Examples of the type of histograms produced to check the data quality are number of tracks reconstructed, energy of calorimeter cells, hit occupancies, synchronization between the different sub-detectors, etc. Some examples are shown in Fig.~\ref{Monitoring}. In the left plot, the occupancy in the SemiConductor Tracker (SCT) modules is shown. The right one shows the difference of the $\theta$ track parameter measured in the inner detector and the muon spectrometer as a function of the event number. It can be seen that both sub-detectors are synchronized.\\
The High Level Trigger algorithms have been providing data streams adequate for different purposes such as alignment and calibrations running at the Cern Analysis Facility (CAF). The first offline data processing takes place with a latency of less than one hour at the Tier-0 and further re-processings with new software versions and updated conditions are done at the Grid Tier-1 centers. At the end of the chain, Event Summary Data (ESD), Monitoring histograms, Combined Ntuples (CBNT), Analysis Data Objects (AOD) and TAG files to allow for a selection of events are produced.\\



\section{Simulation chain}
In addition to the cosmic rays and single beam data taken by ATLAS in the pit, simulated data has also been made available. This is important to check the reconstruction software and allows for data/MC comparisons in terms of detector response, efficiencies, etc. Beam gas and beam halo events have been simulated at 5 TeV. However, as already mentioned, the energy of the LHC proton beam that circulated through ATLAS was of 450 GeV. Consequently, comparisons data/MC have only been made for cosmic rays events.\\
\begin{figure}[!h]
\begin{center}
\includegraphics[width=1.0\textwidth]{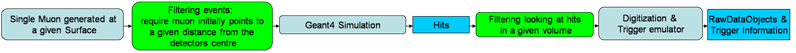} 
\caption{\sl Schematic of the simulation chain.}
\label{Simulation}
\end{center}
\end{figure}
The simulation chain is shown in Fig.~\ref{Simulation}. A Monte Carlo generator was used for simulating muons from cosmic ray events, based on measurements of the differential vertical muon cross section  and analytical calculations extrapolated to low muon energy ~\cite{MC}.  Single muons are generated at the surface.\\
The simulation toolkit Geant4 was used in order to simulate the passage of particles through the detector, giving rise to energy depositions in the detector. In order to reduce the simulation time, only those events pointing to the ATLAS detector are passed to the Geant4 simulation and only those events that have energy depositions in a given volume -in this case, in the Transition Radiation Tracker (TRT) volume- are sent to the next step. This final state consists in emulating the electronics response (the so-called digitization process) in order to end up with simulated raw data.\\

\section{Data Analysis}
\begin{figure}[!h]
\begin{center}
\resizebox{0.6\textwidth}{!}{
\includegraphics[width=0.3\textwidth]{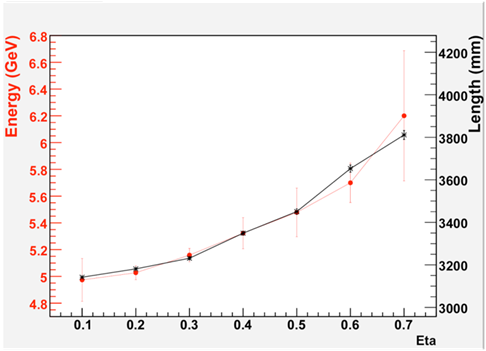}
\includegraphics[width=0.3\textwidth]{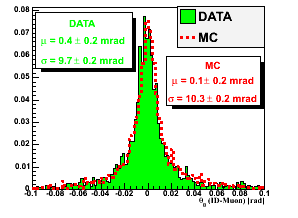}
}\\
\caption{\sl $\eta$ dependence in the hadronic calorimeter response (left). Difference of the $\theta$ parameter measured by the inner detector and muon spectrometer (right).}
\label{DetectorStudies}
\end{center}
\end{figure}

Cosmic rays have allowed us to study the ATLAS detector in terms of efficiencies, resolutions, channel integrity and alignment and calibrations. Some examples of these studies are shown in Fig.~\ref{DetectorStudies}. 
On the left plot, the energy response and track length measured in the hadronic calorimeter (TileCal) is shown. The $\eta$ dependence that is observed is well understood by the variation in the length traversed by the muons in the calorimeter.\\

\begin{figure}[!h]
\begin{center}
\includegraphics[height=0.22\textwidth]{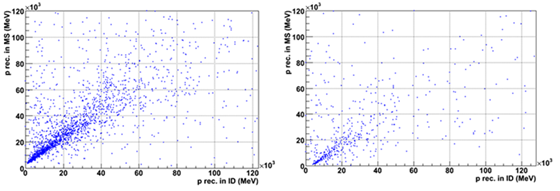}
\caption{\sl Momentum reconstructed in the muon spectrometer and in the inner detector for tracks reconstructed in the top (left plot) and bottom (right plot) part of the muon spectrometer.}
\label{Momentum}
\end{center}
\end{figure}

Furthermore, cosmic ray runs have allowed for studies of the performance of combined algorithms. The ATLAS experiment will identify and measure muons in the muon spectrometer (in a region $|\eta|<$2.7). However, it is not enough. Because of various acceptance gaps in the muon spectrometer and the decrease in the momentum resolution for tracks with low momenta, algorithms that combine the information from the different ATLAS sub-detectors are essential. 
Fig.~\ref{ED2} shows a cosmic ray event in which a combined track (inner detector and muon spectrometer) is reconstructed. The combined tracker matches first tracks of the inner detector with those reconstructed in the muon system and then performs a global $\chi^2$ fit using all hits.\\
The difference between the $\theta$ track parameter measured in these two sub-detectors is shown in the right plot of Fig.~\ref{DetectorStudies}. The mean of these distributions gives an idea of the relative alignment between them. The distributions are more centered for MC, as expected. In addition, the momentum reconstructed in the muon spectrometer and in the inner detector is compared in Fig.~\ref{Momentum} for tracks reconstructed in the top (left plot) and bottom (right plot) part of the muon spectrometer. The difference between the momentum reconstructed in both sub-detectors is the energy deposited in the calorimeters.

\begin{figure}[!h]
\begin{center}
\resizebox{0.5\textwidth}{!}{
\includegraphics[width=0.55\textwidth]{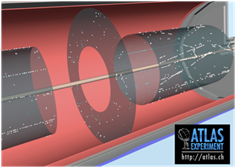}
\includegraphics[height=0.4\textwidth]{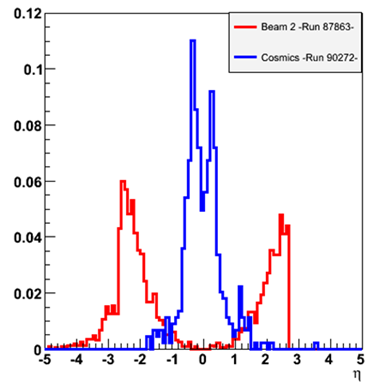}
}\\
\caption{\sl Left: single beam event (halo-like muons). Right: track parameter $\eta$ measured by the inner detector for cosmic rays and single beam events.}
\label{Eta}
\end{center}
\end{figure}

Data taken during the LHC single beam period has also been analysed. Fig.~\ref{Eta} (left) shows a single beam (halo-like muons) event. A comparison of the $\theta$ track parameter measured by the inner detector in cosmic rays and single beam runs is shown in the right plot of Fig.~\ref{Eta}. Cosmics tracks are mainly vertical while single beam tracks are more horizontal.\\

\section{Conclusions}
The complete software chain is being commissioned making use of the different type of data taken by ATLAS, i.e.
cosmic rays and LHC single beam data. All these data have allowed us to study the ATLAS detector in terms of efficiencies, resolutions, channel integrity and alignment and calibration corrections. They have also allowed us to test and optimize the different sub-system reconstruction as well as the muon combined performance algorithms, such as combined tracking tools and different muon identification algorithms based on measurements in the inner detector and muon spectrometer or calorimeters.\\

\begin{acknowledgments}
The commissioning work described here requires many things to be working, from the detectors systems through to the offline software. I would like to acknowledge the huge contribution from the whole ATLAS collaboration.
\end{acknowledgments}

\end{document}